# User Experience of Symptom Checkers: A Systematic Review

Yue You, MS[1], Renkai Ma, MS[1], Xinning Gui, PhD[1]
[1]Pennsylvania State University, University Park, PA, USA

**Abstract**

*This review reports the user experience of symptom checkers, aiming to characterize users studied in the existing literature, identify the aspects of user experience of symptom checkers that have been studied, and offer design suggestions. Our literature search resulted in 31 publications. We found that (1) most symptom checker users are relatively young; (2) eight relevant aspects of user experience have been explored, including motivation, trust, acceptability, satisfaction, accuracy, usability, safety/security, and functionality; (3) future symptom checkers should improve their accuracy, safety, and usability. Although many facets of user experience have been explored, methodological challenges exist and some important aspects of user experience remain understudied. Further research should be conducted to explore users' needs and the context of use. More qualitative and mixed-method studies are needed to understand actual users' experiences in the future.*

**Introduction**

The rapid development of digital technology and the healthcare industry has contributed to the increased popularity of symptom checkers (SCs) in application (app) markets. Some SC apps (e.g., Ada) have been downloaded from app stores tens of millions of times[1]. These SCs utilize either a chatbot or a questionnaire-like form to solicit symptom input and provide users with a potential diagnosis. Some SCs describe themselves as akin to intelligent doctors who give accurate and reliable medical recommendations. For instance, Ada claims to "*think like a doctor*"[2]. However, the reliability of SCs is questionable[3]. Healthcare consumers may put their health at risk if they uncritically accept the diagnoses from SCs[4]. Recognizing that individuals' perceptions and interpretations of SCs' responses may influence their subsequent actions, it is important to understand more fully the user experience (UX) of SCs[5]. UX is "a consequence of a user's internal state, the characteristics of the designed system, and the context (or the environment) within which the interaction occurs.[6]" As Table 1 indicates, we adapted this definition of UX to characterize the UX of SCs along three dimensions. As the UX of SCs is highly related to the acceptance of SCs[7] and the minimization of unintended consequences (e.g., anxiety and cyberchondria)[8], it is crucial to improve the UX of SCs by understanding individuals' motivations for using SCs, users' expectations of SCs, the benefits as well as the limitations of SCs' features, and the usage contexts.

**Table 1.** Three dimensions of the UX of SCs.

| Dimension | Definition |
|---|---|
| Consequence of a user's internal state | User's expectations of, needs for, and attitude towards SCs |
| Characteristics of the designed system | SCs' features, including usability, complexity, purpose, and functionality |
| Context or environment | Context or environment within which users interact with SCs |

Nevertheless, few existing papers have systematically reviewed studies of the UX of SCs. Instead, previous systematic reviews of SCs have focused on the influence of SCs' use on the patient–doctor relationship[9], the current evaluation methodologies for SCs[1], and the impact of SCs on urgent care seeking[10] and primary care[11]. To the best of our best knowledge, only one scoping review of SCs covered studies addressing the UX of SCs[12]. This scoping review provided a broad view of research on SCs and points out that UX is one theme that has been explored. However, it did not offer an in-depth analysis of the aspects of the UX of SCs that have been explored. It is still unclear who SC users are, what factors motivate and influence the use of SCs, and what relevant design suggestions might be. To address this gap in existing reviews, we systematically reviewed papers that examine the UX of SCs. We choose the systematic review approach instead of doing a scoping review because a systematic review can allow us to synthesize the findings from previous literature and identify areas for future research[13]. We intended to address the following three questions: (1) Who uses SCs? (2) Which aspects of the UX of SCs have been studied based on the three dimensions of UX? and (3) What design suggestions have been proposed to improve the UX of SCs?

**Materials and Methods**

*Search strategy and screening process.* We developed a study protocol in compliance with the PRISMA guidelines. The PROSPERO website was searched and no previous systematic review on similar topics was found, which validated the novelty of our review. We searched six electronic databases from May 2021 to July 2021: PubMed,

CINAHL, ACM Digital Library, Scopus, Web of Science, and Google Scholar. We also conducted backward–forward searches by examining the references from relevant articles and systematic reviews. We utilized three groups of search terms: SC-related terms, internet technology-related terms, and health-related terms. The following search query was used for Scopus: (TITLE-ABS-KEY("symptom checker" OR "self diagnosis" OR "self-diagnosis" OR "self triage" OR "self-assessment" OR "self assessment" OR "self diagnosing" OR "self-evaluation" OR "self evaluation" OR "self-appraisal" OR "check your symptom" OR "check their symptom" OR "online diagnosis" OR "web-based triage" OR "electronic triage" OR "etriage" OR "self referral" OR "assistance technology" OR "assistance technologies") AND TITLE-ABS-KEY("online" OR "technology" OR "website" OR "tool" OR "computer" OR "mhealth" OR "m-health" OR "ehealth" OR " e-health" OR "app" OR "mobile application" OR "smartphone" OR "smart phone" OR "cell phone" OR "cellular phone" OR "mobile phone" OR "electronic" OR "automated" OR "internet" OR "digital" OR "mobile") AND TITLE-ABS-KEY("medical" OR "health" OR "medication" OR "healthcare")). The search strings for Scopus were adapted for other databases.

After completing the database and backward–forward searches, we conducted two rounds of screening to determine which articles to include in our review. First, we screened the titles and abstracts of the identified papers using three eligibility criteria: (1) the paper is an original research article published in a peer-reviewed journal or conference proceedings; (2) the paper was published in English; and (3) the paper's title and abstract included our search terms. We then screened the full texts of those papers based on the following inclusion criteria: (1) the subject of the paper is an application rather than a model or algorithm or general online search; (2) consumers/laypersons are the primary users of the technology developed and/or evaluated in the paper; (3) the technology described in the paper is internet-enabled; (4) the technology studied in the paper is used for self-diagnosis; (5) the topic of the paper addresses an aspect of the UX of SCs; and (6) the full text is available. After each round of screening, three researchers performed a cross-review and settled any disagreements. Only those papers that the researchers deemed sufficiently relevant were included. In addition, quality assessment was conducted for all papers to assess whether these papers' aims and methodology were clearly described, whether their findings align with their research questions, and whether they were peer-reviewed. All discrepancies were discussed and resolved. The quality assessment results were taken into account when synthesizing data. We finally identified 16,825 potentially relevant articles written in the English language. We also identified 3,425 papers using Google Scholar and backward–forward searching. Ultimately, 31 papers (24 from databases and 7 from other methods) were chosen to be included. Figure 1 shows the screening process.

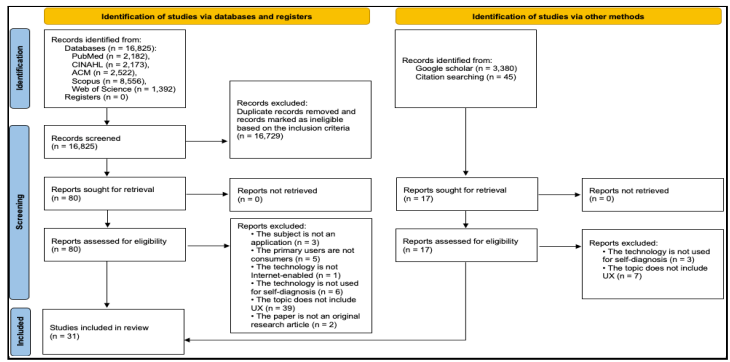

**Figure 1.** Processes of identification and screening.

*Data collection and analysis.* Using the papers identified during the two rounds of screening, we conducted data extraction based on the PRISMA guidelines. We collected the following main information: (1) the source of the paper, including the author(s), date of publication, and publication venue; (2) background information, including the diseases examined in the paper, the research purposes, stakeholders, and the features/functions of SCs; (3) methodological information, including each author's country and affiliation, the study setting and duration, the participants' characteristics, the data collection and analysis methods; and (4) findings, including the participants' motivations for using SCs, the participants' characteristics, the dimensions/measurements of UX, and design suggestions. We then used thematic analysis and descriptive statistical analysis to synthesize the extracted data.

**Results**

*Study characteristics.* The 31 papers chosen for inclusion were published between 2007 and 2021 (see Table 2).

**Table 2.** List of studies included in the systematic review.

| Source | Country | Study Design | Methodology | Diseases | Participants | System Features |
|---|---|---|---|---|---|---|
| Powley et al, 2016[14] | UK | Laboratories study | Quantitative; Experimental study | Inflammatory joint disease | N = 34; Age: 29.75-78.5; Female | Existing apps; Questionnaire-based |
| Bisson et al, 2016[15] | US | Laboratories study | Quantitative; Experimental study | Knee pain | N = 328; Age: 18-76; 49.7% male | Prototype; Questionnaire-based |
| Bisson et al, 2014[16] | US | Field study | Quantitative; Experimental study | Knee pain | N = 527; Age: avg 47, 18-84; 51.6% male | Prototype |
| Hageman et al, 2015[17] | US | Laboratories study | Quantitative; Experimental study & questionnaire | Hand diseases | N = 86 patients; Avg age: 46; 51% male | Existing apps; Questionnaire-based |
| Knitza et al, 2021[18] | Germany | Laboratories study | Quantitative; Experimental study | Inflammatory rheumatic disease | N = 164 patients | Existing apps; Chatbot & questionnaire-based; AI-driven & rule-based |
| Cross et al, 2021[19] | Australia | Laboratories study | Quantitative; Experimental study & questionnaire & log-based study | N/A | N = 64; Age: > 18, 87.49%: 18-34; 4.68%: 35-44; 1.56%: 45-54 | A clone of existing app; Questionnaire-based |
| Verzantvoort et al, 2018[20] | the Netherlands | Field study | Mixed methods; Experimental study & interview | Acute primary care | N = 4,456; Age: most 19-45 (56.4%) | Existing app |
| Fan et al, 2021[21] | China | Field study | Mixed methods; Log-based study | N/A | N = 16,519; Age: most 20-39; 54.8% male | Existing app; Chatbot; AI-driven |
| Price et al, 2013[22] | US | Laboratories study | Quantitative; Experimental study & questionnaire | Influenza-Like Illness | N = 294 parents and caregivers; Age: > 18 | Prototype; Questionnaire-based |
| You & Gui, 2020[23] | US | Field study | Qualitative; App review analysis & interview | N/A | N = 10 (interview); Age: 23-40 (interview) | Existing apps; Chatbot |
| Aboueid et al, 2021[7] | Canada | Laboratories study | Qualitative; Think aloud & interview | N/A | N = 24; Age: 18-34 | Existing apps; Questionnaire-based & chatbot |
| Aboueid et al, 2021[24] | Canada | Laboratories study | Qualitative; Interview | COVID-19 | N = 22; Age: 18–33; 54.5% female | Existing apps; Questionnaire-based & chatbot |
| Marco-Ruiz et al, 2017[25] | Norway | Laboratories studies | Mixed methods; Experimental study & think aloud | Respiratory diseases | Phase 1: N = 53; Phase 2: N = 15 | Existing apps |
| Ponnada, 2020[26] | US | Laboratories study | Mixed methods; Usability testing & interview | COVID-19 | N/A | Existing app & improved prototype; Chatbot |
| Schrager et al, 2020[27] | US | Laboratories study | Quantitative; Prototyping & usability testing | COVID-19 | N = 877; Age: avg. 32, 19-84; 65.3% female | Prototype; Questionnaire-based |
| Reilly & Austin, 2021[28] | Australia | Field study | Quantitative; Online survey | Depression in pregnancy | N = 140; Age: N/A; Female | Existing app; Questionnaire-based |
| Nieroda et al, 2018[29] | UK | Laboratories study | Qualitative; Think aloud & online survey | Cancer | The public: N=39, Age: >= 40; Practitioners: N = 20 | Existing app |
| Daher et al, 2020[30] | Switzerland | Laboratories study | Quantitative; Experimental study | N/A | N = 12; Age: N/A | Prototype; Chatbot |
| Tsai et al, 2021[31] | US | Field study & laboratories study | Mixed methods; Interview & experimental study | General/ COVID-19 | (1) N = 25; Age: 19-54;(2): N = 20; Age: 19-37; 70% female | Existing apps & prototype; Chatbot & questionnaire-based |
| Li et al, 2020[32] | China | Field study & laboratories study | Mixed methods; Interview & experimental study | N/A | (1) N = 13; Age: 21-61 (2) N = 48 | Existing apps; Chatbot; AI-driven |
| You et al, 2021[33] | China | Field study | Qualitative; Interview | N/A | N = 30; Age: 20-55 Most were in their 20s and 30s | Existing apps; Chatbot & questionnaire-based; AI-driven |

**Table 2.** List of studies included in the systematic review (continued).

| Source | Country | Study Design | Methodology | Diseases | Participants | System Features |
|---|---|---|---|---|---|---|
| Hwang et al, 2018[34] | US, South Korea | Laboratories study | Qualitative; Think aloud | N/A | N = 6; Age: 20-39; 4 female, 2 male | Existing apps; Chatbot; AI-driven |
| Baldauf et al, 2020[35] | Switzerland | Questionnaires | Mixed methods; Online survey | N/A | N = 106; Age: avg: 34, 19-60; 58.5% male | Existing apps; Using sensor data; AI-driven |
| Miller et al, 2020[36] | UK | Laboratories study | Quantitative; Experimental study & questionnaire | N/A | N = 523; Age: avg: 39.79, 81.6%: 15-64; 62.1% female | Existing app Chatbot; AI-driven |
| Meyer et al, 2020[37] | US | Questionnaires | Mixed methods; Online survey | N/A | N = 329; Age: avg: 48.0; 75.7% female | Existing app; questionnaire, AI-based |
| Winn et al, 2019[38] | US | Field study | Quantitative; Log analysis | N/A | N = 158,083; Age: avg: 40; 78% female | Existing app; Chatbot |
| DeForte et al, 2020[39] | US | Field study | Qualitative; App review analysis | Depression | N/A | Existing apps |
| Nijland et al, 2010[40] | the Netherlands | Field study & questionnaires | Quantitative; Log-based & survey | N/A | N = 192; Age: 16–35 (56%); 66% female | Existing app; Questionnaire-based |
| Luger et al, 2014[41] | US | Laboratories study | Qualitative; Think aloud & questionnaire | N/A | N = 79; Age: >= 50; | Existing app; Questionnaire-based |
| Lanseng et al, 2007[42] | Norway | Questionnaires | Quantitative; Online survey | N/A | N = 160; Age: 18-65; 46.3% female | N/A |
| Hua & Hou, 2020[43] | Vietnam | Field study & questionnaires | Mixed methods; Interview & survey | N/A | N = 482; Age: 18-50; 63.1% female | N/A |

*Characteristics of users.* Four studies that we reviewed (12.9%)[7,21,37,38] reported on the characteristics of individuals who had used SCs prior to participating in the studies. The age of SC users was found to be relatively young (most users ranged in age from 20 to 39 years old[21] or the average age was 40 years old[38]). Regarding the gender of SC users, the study conducted in China found male users initiated most consultations (54.80%)[21], while the U.S. study reported most users were female (75.7%)[37]. Other less commonly mentioned user characteristics in our corpus included health literacy[7], education[37], and income[37]. Young individuals with low health literacy and high technology literacy were found to be inclined to use SCs[7]. The household income of most users was less than $100,000 USD[37].

*UX aspects of SCs.*

(1) Consequence of a consumer's internal state
**Motivations for using SCs**: Five studies (16.1%)[21,28,33,38,43] addressed users' motivations for using SCs (see Table 3).

**Table 3.** Motivations for using SCs (n = 5).

| Motivation | Details |
|---|---|
| Convenience | The researchers found that individuals were motivated to use SCs instead of going to a doctor because of the time and cost savings as well as location independence[33] |
| Supplementing offline medical visits | Individuals were motivated to use SCs to supplement in-person medical visits and gain assistance in making healthcare-related decisions[28,33,38] |
| Gaming | Individuals (8.03%) gamed SC by entering nonsensical words for nontherapeutic purposes[21,43] |
| Social influence | Social influence (i.e., "the extent to which an individual is aware that significant others recommend they use the self-diagnostic app") could prompt consumers to use SCs[43] |

**Trust**: Six of the reviewed papers (19.4%) explored users' trust in SCs[21,31–35], reporting factors that may influence users' trust, including the provision of explanations, interaction design, and accuracy (see Table 4).

**Table 4.** Factors that influence the users' trust (n = 6).

| Factor | Impact | Methods |
|---|---|---|
| Explanations | The information source or authority[31,33], medical certification[35], security of data transmission[35], and explanations for presented questions and reports[31,34] could positively influence users' trust | Interviews[33], think aloud[34], questionnaires (adapted from previous literature[31], self-created[35]) |
| Interaction design | Repeated or irrelevant questions[34], vague options[34], limited input functions[33], and an overly anthropomorphic conversation design[33] could negatively affect users' trust in SCs | Interviews[33], think aloud[34] |
| Accuracy | Users tended to distrust SCs when the SCs' recommendations were inaccurate[21,32,33] | Interviews[33], log analysis[21], questionnaires (self-created)[32] |

**Acceptability**: In our corpus, a total of four papers (12.9%) measured whether users accepted SCs[18,28,29,36]. These studies relied on questionnaires[18,28,29,36] and the thinking-aloud method[29]. All of the papers described how most users accepted the use of SCs to support emotional wellness[28], primary care[36], and cancer diagnosis[29].

**Satisfaction**: Four of the studies (12.9%) evaluated individuals' satisfaction with using SCs via questionnaires[19–21,31]. These studies found that some of the relevant factors affecting satisfaction are disease type[21], accuracy[21], and explanations[31]. One log-based study found that for common diseases, users were usually satisfied with the diagnostic recommendations, while for more complex diseases, users were less satisfied[21]. This study also reported that inaccurate diagnostic results contributed to dissatisfaction with SCs. One experimental study reported that providing different styles of explanations can increase users' satisfaction with SCs[31].

(2) Characteristics of the SCs

**Accuracy**: Eleven papers included in our corpus examined the accuracy of SCs. Three papers stated that compared with physicians' diagnoses, recommendations from SCs were relatively accurate in diagnosing knee pain (correctly diagnose 58% of the time)[15], acute primary care (correctly triage 89% of patients)[20], and influenza-like illness (correctly diagnose 93% of the time)[22], though two of them admitted that SCs had low specificity[16,22]. However, the other eight papers revealed that SCs are frequently inaccurate when providing diagnoses for inflammatory joint diseases[14], hand diseases[17], inflammatory rheumatic diseases[18], knee pain[15], and general diseases (e.g., 1,084 out of 3,832 diagnoses were rated by users as inaccurate[21])[19,21,37,41].

**Usability**: Ten papers explored the usability of SCs. Three experiment-based papers reported that SCs were easy to use[18,22,36] (e.g., 97.8% of the participants considered Ada easy to use[36]) by using the standardized scales (e.g., System Usability Scale)[18,36] or self-created web survey[22]. The remaining seven studies reported usability issues (see Table 5).

**Table 5.** Usability issues (n = 10).

| Usability issues | Details | Methods |
|---|---|---|
| Neglect of user groups and diseases | SCs could not support diverse user groups (e.g., transgender individuals) or diseases[23] | Interviews[23], app review analysis[23] |
| Input limitations | Users perceived difficulties in inputting the information of their diseases completely (62%)[20], inputting sufficient patient history information[23], describing disease dimensions (e.g., frequency)[23,44], and entering medical terms that could be recognized by the SCs[23] | Questionnaires (self-created)[20], think aloud[44], interviews[23], app review analysis[23] |
| Problematic questions | Users perceived that the probing questions of SCs were seemingly irrelevant based and were presented in an unreasonable sequence[23] | Interviews[23], app review analysis[23] |
| Interpretation issues | Users encountered challenges in understanding medical jargon[7], complex questions[23], and diagnostic reports[21,31] | Interviews[7,23], app review analysis[23], log analysis[21], questionnaires (adapted from previous literature)[31] |
| Overwhelming amounts of information | Users complained of too much information on the welcome page[44], excessive subsections to fill in[44], too many pieces of information presented at once[26], and redundant questions[23] | App review analysis[23], interviews[23,26], usability testing[26], think aloud[44] |

**Safety/security**: Three studies highlighted that using SCs may have potential risks for both the healthcare system and consumers[22,28,33]. For instance, the specificity of the SC used for influenza-like illness is poor, which may lead more users to go to the emergency department, thereby putting pressure on healthcare systems[22]. Likewise, since most health consumers are laypersons who may not have sufficient medical knowledge, they cannot discern the correctness of SCs' recommendations. Using SCs uncritically may compromise users' health and safety[28,33]. For example, one paper found only 53% to 61% of users with depression reported by the focal SC discussed their depression with medical professionals[28]. This might result in untreated depression. Two papers emphasized the significance of data security[35,44], calling for more information about how SCs encrypt[35], anonymize[35], and store data[44].

**Functionality**: Two studies examined the functions of SCs using an app feature analysis[23], an app review analysis[23,39], and semi-structured interviews[23]. The first study examined 11 apps released in the United States via the Apple App store and the Google Play store[23]. The researchers found most apps support five procedures: creating a patient history for users, evaluating symptom input by users, providing recommendations, ordering further tests, and providing follow-up treatment (e.g., contacting healthcare practitioners). The other study showcased the multiple functions users desired, such as mood tracking, journaling, and providing educational materials[39].

(3) Context

Four studies[20,21,38,40] mentioned the context within which consumers interacted with SCs. They reported that SCs were used to address myriad symptoms. Two of them reported users mostly used SCs to address non-emergency care issues, such as common cold symptoms[40], gastroenterology[21], and dermatology[21], while fewer individuals used SCs in emergent situations[21]. The SC DoctorBot was also used for diseases that may carry social stigma[21]. One of the studies found that the most common symptoms reported by users are pain and abdominal functioning[38].

*Design suggestions for improving the UX of SCs.* Based on their findings, the papers included in our corpus offered design recommendations for improving the UX of SCs. The details are presented in Table 6.

**Table 6.** Design suggestions.

| Aspects | Design suggestions |
| --- | --- |
| Accuracy | Increase the accuracy of SCs by considering the probability of specific conditions[17] and the uniqueness of a variety of symptom clusters[17] |
| Safety | • Supply extra healthcare information[20] and provide guidance as well as warnings to the public[29,33]<br>• Enact relevant policies or legislation to regulate online SC consultation, such as asking SC suppliers to clarify the responsibility of SCs[40], pre-test SCs[42], and develop a consumer protection policy[43] |
| Functions | • Provide functions that are similar to those of in-person medical visits (e.g., offering customized symptoms tracking, contacting medical professionals, and providing medications information)[23,33]<br>• Provide a comprehensive health profile and follow-up treatment[23] |
| Broader healthcare information | • Introduction to SC functions, the consultation process[21], and prediction accuracy[21], information about diagnostic conditions[35], Treatment options[21,35]<br>• Endorsement of established authorities[33]<br>• Explanation of data sources[21,31], probing questions[31,34], and the decision model[21,31,34]<br>• Treatment options[21,35] and links to other information sources[21] |
| Usability | • Have flexible input functions (e.g., voice recordings[21] and approximate string matching function[23])<br>• Convey human empathy[26,30,32] and use layperson language[23,29], and provide a good information architecture[26] (i.e., present key information logically instead of showing multiple pieces of complex information at once)<br>• Use readable fonts as well as vivid color[26], visual summaries (e.g., tables and figures)[41], and navigation aids[41] |

**Discussion**

This systematic review is the first to address the UX of SCs. We reported on and analyzed the characteristics of the published literature, SC users' characteristics, the motivations of individuals to use SCs, the relevant aspects of the UX of SCs, and design implications. In this section, we discuss the implications of our results for future research.

Overall, the UX studies in our corpus focus primarily on the characteristics of the SCs rather than investigating consumers' internal states. The included studies put more emphasis on evaluating rather than designing SCs centering around users' needs. In healthcare, patient-centered care (i.e., "providing care that is respectful of and responsive to individual patient preferences, needs, and values"[45]) is always highlighted. When designing healthcare technologies, users' expectations and needs should also be centered and considered[46]. However, insufficient attention has been paid to fully understanding users' needs for SCs, especially their concerns and expectations in everyday practices. Our findings show that most studies have explored the use of SCs in laboratory or otherwise controlled settings (n=16). While we acknowledge the importance of laboratory experiments, we suggest that studies that are deployed in the wild are needed to capture individuals' unanticipated use of SCs and reveal UX issues that only arise in-situ[47].

In addition, patient-centered care suggests giving respect, showing emotional support, communication skills, and education, and offering explanations[48,49]. However, whether and how this line of guidelines should be applied to SC design is still unclear. First, our systematic review found that few studies focus on how to provide emotional support to SC users and improve their hedonic experience. Only four studies highlighted that SCs should convey human empathy[26,30,32]. Little is known about how to deliver emotional support to users. Moreover, although existing healthcare research suggests that humanoid features of healthcare technologies may improve communication and therefore UX[50,51], little research in our corpus put forward recommendations for how to embed humanoid features in SCs to improve the hedonic experience. Further research should be conducted to investigate what level of empathy should be provided and how the interface design of SCs can be improved. Second, though four studies we reviewed investigated the explanations[21,31,34,35], which emphasized the importance of providing explanations (e.g., explanations for presented questions and diagnostic reports[31,34]) to improve the transparency of SCs and users' trust, it is still unclear how we should deliver explanations to users. Future research might explore how to give users more control and customization over the offerings of explanations[31].

Furthermore, few studies in our corpus focus on the context of use. Existing research has highlighted the significance of the context in which technology is embedded[52], recognizing that technology is integrated into everyday practices[52]. Health information needs are especially likely to vary across locations[53]. It is crucial to consider the geographic and other situated contexts within which the user interactions with SC happen. For instance, our review shows that most papers on SCs are Western in origin and their studies were conducted in developed countries. Few studies were conducted in developing countries such as Vietnam (n = 1). Since healthcare systems vary across geographic contexts (from country to country[54,55] and from rural to urban areas[56]), it is possible that the findings of our reviewed papers would change if these studies were conducted in new geographic contexts. Our review indicates that future research should be including more diverse geographic contexts, such as undeveloped and developing countries.

*Implications for future SC research.* **More qualitative and mixed-methods research is needed.** A key takeaway of our review is that the majority of papers used quantitative methods (n = 14). These studies measured UX via questionnaires or through clinical vignettes. Few of the studies we reviewed investigated in depth why individuals decide to use, trust, and are satisfied with SCs. We call for more qualitative research exploring the deeper reasons behind users' preferences. In addition, mixed methods are common in healthcare research as they can improve the research rigor and offer a more comprehensive portrayal of the studied phenomenon[57,58]. Though the qualitative studies and mixed-method studies in our corpus revealed how individuals' trust in SCs, SCs' usability, and SCs' accuracy were measured, there were only a few such studies. To validate the qualitative data and produce more generalizable results, future research needs to complement the qualitative data with quantitative data from large samples[59].

**More diverse user groups should be explored.** Our review suggests that future UX research should consider diverse user groups. One paper we reviewed argued that when designing new SCs, programmers should consider a diversity of user groups, as existing SCs lack support for special user groups, such as transgender individuals[23]. This is in line with the principles of inclusive design. Inclusive design calls for the consideration of all users' needs without adaptation[60]. Prior research has highlighted the importance of inclusiveness in healthcare services[61]. The reviewed studies reported on some user characteristics, such as age, gender, education level, health literacy, and household income. However, the number of studies we captured is quite limited (n = 4), and there is a lack of research on how users with different characteristics (e.g., different age, gender, and health literacy levels) experience SCs. When developing healthcare apps, all health literacy levels should be thoroughly considered since individuals with different health literacy levels may have different needs[62]. For example, older adults may have more barriers to using healthcare technologies due to limited health and technology literacy[63]. The gender differences may lead to different acceptance of SCs as people of different genders may have distinct attitudes toward healthcare applications[64]. It is thus important to uncover the diverse needs and experiences of users of different age, gender, and health literacy levels. In the future, researchers might look into how special user groups experience SCs.

**More focus on safety and ethics is needed.** In healthcare, safety and security problems need to be prioritized since user information is sensitive[65]. Recent research on the reliability and accountability of health applications has pointed out that there are no clear measures to assess and certify SCs and guarantee their reliability and quality[66], which may lead to medical errors and cause unintended harm to users due to the nature of smartphones[67]. Several papers in our review also highlighted potential risks of using SCs, including poor specificity[22] and data security issues[35,44]. However, there were few suggestions regarding how to ensure the safety of current SCs. To deliver safe healthcare services, it is also necessary to consider the ethical challenges or the moral considerations[68]. As noted in the included studies[7,35], users were always concerned about their health data privacy. However, ethical issues regarding SC's anonymity and data collection/storage were not considered in the included studies. Few studies in our corpus offered recommendations for solving ethical challenges and enacting new policies, such as establishing a data protection policy[43] or testing SCs before release[7,39]. It is still unclear whether the developers of SC apps have sufficient credentials, disclosed enough information to ensure transparency, and guaranteed data security. Furthermore, previous research has pointed out that AI systems need regulation and strict testing before being released to the public[69] since AI technologies may introduce risks due to their opaque algorithms[70]. Although eight papers in the corpus described their focal SCs as AI-driven, few put forward policies to ensure users' safety. Thus, future research should explore the kinds of policies that should be made to govern SCs, addressing SCs' accountability, data privacy and security.

**Limitations of the review process.** The present paper has several methodological shortcomings. First, papers not written in English were excluded. Second, due to resource constraints, we may have missed relevant contemporary papers spurred by the fast-moving development of SCs. In fact, the ACM Digital Library's search engine limits database users to the first 2,000 papers that appear in the search engine's index, even if they modify their search keywords several rounds to acquire as many as possible articles as we did. To limit the risk of missing papers, we searched multiple databases as well as Google Scholar and conducted a backward–forward reference check.

**Conclusion**

This systematic review summarizes trends in the UX of SCs. Despite the limitations, our review contributes to the SC research by providing a thorough overview of prior UX research and discussing factors that influence SC usage. Our review presents SC users' characteristics, individuals' motivations to use SCs, aspects of the UX of SCs, and design implications for SCs. More user studies are required to narrow the research gaps identified in our study.